\documentclass{ws-ijmpa}

\begin{document}

\markboth{F. Huang, Z.Y. Zhang and Y.W. Yu}{$S$, $P$, $D$, $F$
wave $KN$ phase shifts in the chiral SU(3) quark model}

\catchline{}{}{}{}{}

\title{$S$, $P$, $D$, $F$ wave $KN$ phase shifts in the chiral SU(3) quark model}

\author{F. Huang}
\address{Institute of High Energy Physics, P.O. Box 918-4,
Beijing 100049, PR China \\ Graduate School of the Chinese Academy
of Sciences, Beijing 100049, PR China}

\author{Z.Y. Zhang and Y.W. Yu}
\address{Institute of High Energy Physics, P.O. Box 918-4,
Beijing 100049, PR China}

\maketitle


\begin{abstract}
The $S$, $P$, $D$, $F$ wave $KN$ phase shifts have been studied in
the chiral SU(3) quark model by solving a resonating group method
equation. The numerical results of different partial waves are in
agreement with the experimental data except for the cases of
$P_{13}$ and $D_{15}$, which are less well described when the
laboratory momentum of the kaon meson is greater than 400 MeV.
\end{abstract}

\keywords{$KN$ phase shifts; quark model; chiral symmetry.}

\section{Introduction}

The kaon-nucleon ($KN$) scattering processes have aroused
particular interest in the past
\cite{rbu90,nbl02,dha02,sle03,hjw03} and many works have been
devoted to this issue based on the constituent quark model. But up
to now, most of them cannot accurately reproduce the $KN$ phase
shifts up to the orbit angular momentum $L=3$ in a sufficient
consistent way. In this work we perform a resonating group method
(RGM) calculation of $S$, $P$, $D$, $F$ wave $KN$ phase shifts of
isospin $I=0$ and $I=1$ in the chiral SU(3) quark model, which has
been successful in reproducing the energies of the baryon ground
states, the binding energy of deuteron, the nucleon-nucleon ($NN$)
scattering phase shifts, and the hyperon-nucleon ($YN$) cross
sections by the RGM calculations \cite{zyz97,lrd03}. The model is
extended to include an antiquark $\bar s$ and the mixing of
$\sigma_0$ and $\sigma_8$ is considered. A satisfactory
description of the $KN$ phase shifts for different partial waves
is obtained except for the cases of $P_{13}$ and $D_{15}$, of
which the calculated phase shifts are too repulsive and a little
bit too attractive respectively when the laboratory momentum of
the kaon meson is greater than 400 MeV in this present
investigation.

\section{Formulation}

In the chiral SU(3) quark model, the potential between the $i$th
and $j$th constituent quarks can be written as
\begin{equation}
V_{ij}=\sum_{i<j}(V^{conf}_{ij}+V^{OGE}_{ij}+V^{ch}_{ij}),
\end{equation}
where the confinement potential $V^{conf}_{ij}$ describes the
long-range nonperturbative QCD effect and the one-gluon-exchange
potential $V^{OGE}_{ij}$ depicts the short-range perturbative QCD
effect. The chiral-field-induced quark-quark potential is in the
form of
\begin{equation}
V^{ch}_{ij}=\sum^{8}_{a=0}V_{\sigma_a}(\bm
r_{ij})+\sum^{8}_{a=0}V_{\pi_a} (\bm r_{ij}),
\end{equation}
and mainly signifies the medium-range nonperturbative QCD effect.
In this expressions, $\sigma_{0},...,\sigma_{8}$ are the scalar
nonet fields, and $\pi_{0},...,\pi_{8}$ the pseudoscalar nonet
fields. In order to study the $KN$ system, we extend our model to
include an antiquark $\bar s$. The interaction between $u(d)$ and
$\bar s$ includes two parts: direct interaction and annihilation
parts,
\begin{equation}
V_{i\bar 5}=V^{dir}_{i\bar 5}+V^{ann}_{i\bar 5},
\end{equation}
where
\begin{equation}
V_{i\bar 5}^{dir}=V_{i\bar 5}^{conf}+V_{i\bar 5}^{OGE}+V_{i\bar
5}^{ch},
\end{equation}
\begin{equation}
V_{i\bar 5}^{ann}=V_{ann}^{K}.
\end{equation}
Now, the total Hamiltonian of $KN$ system is written as
\begin{equation}
\label{hami5q}
H=\sum_{i=1}^{5}T_{i}-T_{G}+\sum_{i<j=1}^{4}V_{ij}+\sum_{i=1}^{4}V_{i\bar
5},
\end{equation}
where $T_G$ is the kinetic energy operator of the center of mass
motion, and the explicit expressions of the potentials can be
found in the literature \cite{fhu03,fhu41,fhu42}.

In our calculation, the mixing of $\sigma_0$ and $\sigma_8$ is
considered, and the mixing angle $\theta^S$ is taken to be two
possible values. One is $35^\circ$ (ideal mixing) and the other is
$-18^\circ$ (provided by Dai {\it et al.} \cite{ybd03}). The model
parameters are fixed by some special constraints
\cite{fhu41,fhu42} and their values are tabulated in Table 1.
\begin{table}[h]
\tbl{Model parameters. The meson masses are taken to be the
experimental data except for $m_\sigma$ which is taken to be 675
MeV. The cutoff mass $\Lambda=1100$ MeV.}
{\begin{tabular}{@{}cccccccccc@{}} \toprule
$\theta^S$ & $b_u$ & $m_u$ & $m_s$ & $g_u$ & $g_s$ & $a_{uu}^c$ & $a_{us}^c$ & $a_{uu}^{c0}$ & $a_{us}^{c0}$\\
           & (fm)  & (MeV) & (MeV) &       &       &(MeV/fm$^2$)&(MeV/fm$^2$)&   (MeV)       &   (MeV)      \\ \colrule
$35^\circ$ & 0.5   &  313  &  470  & 0.886 & 0.917 &    52.4    &    72.3    &   -50.4       &   -54.2      \\
-$18^\circ$& 0.5   &  313  &  470  & 0.886 & 0.917 &    55.2    &    68.4    &   -55.1       &   -48.7      \\ \botrule
\end{tabular}}
\end{table}

\section{Results and discussions}

\begin{figure}[h]
\begin{center}
\psfig{file=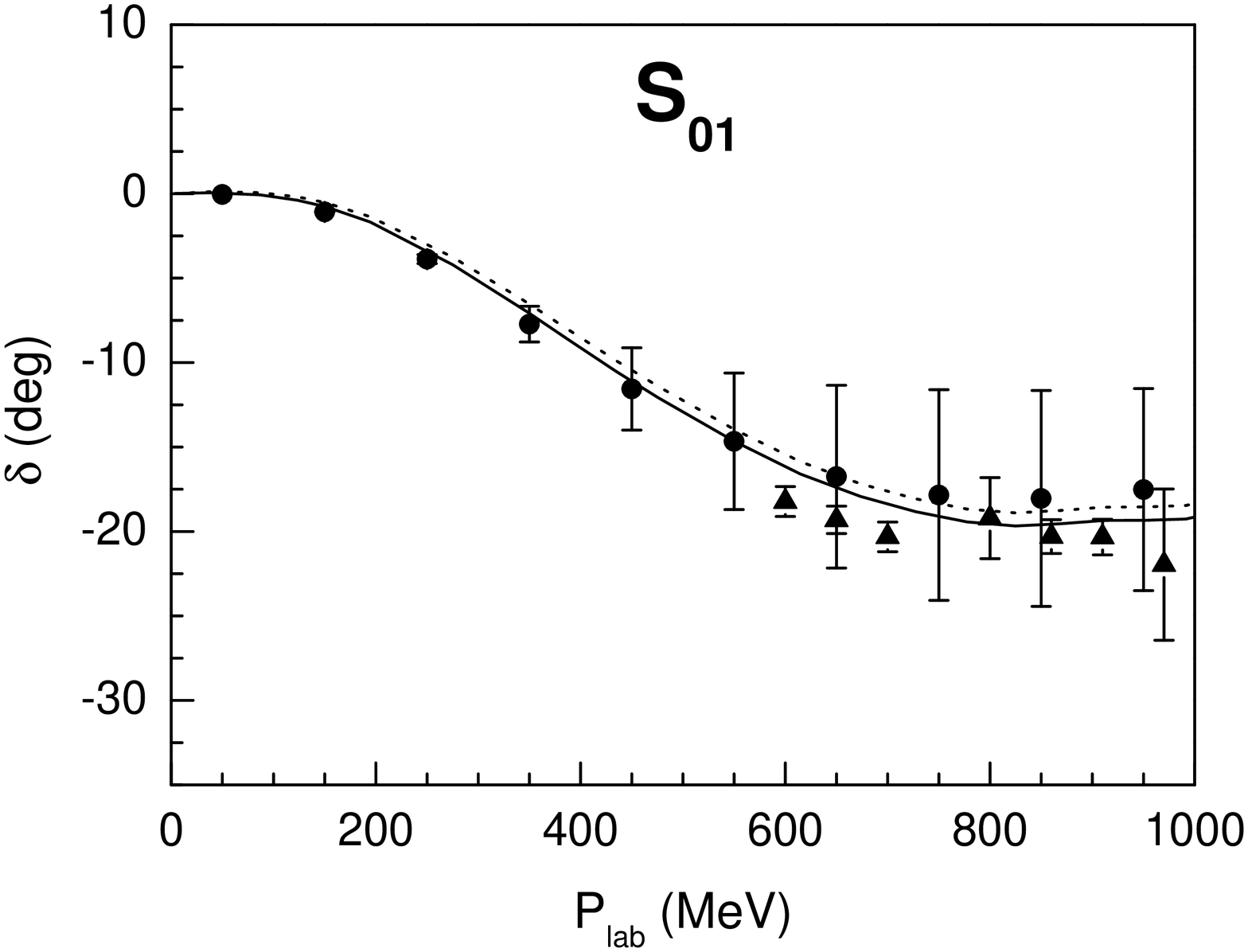,width=4.45cm}
\psfig{file=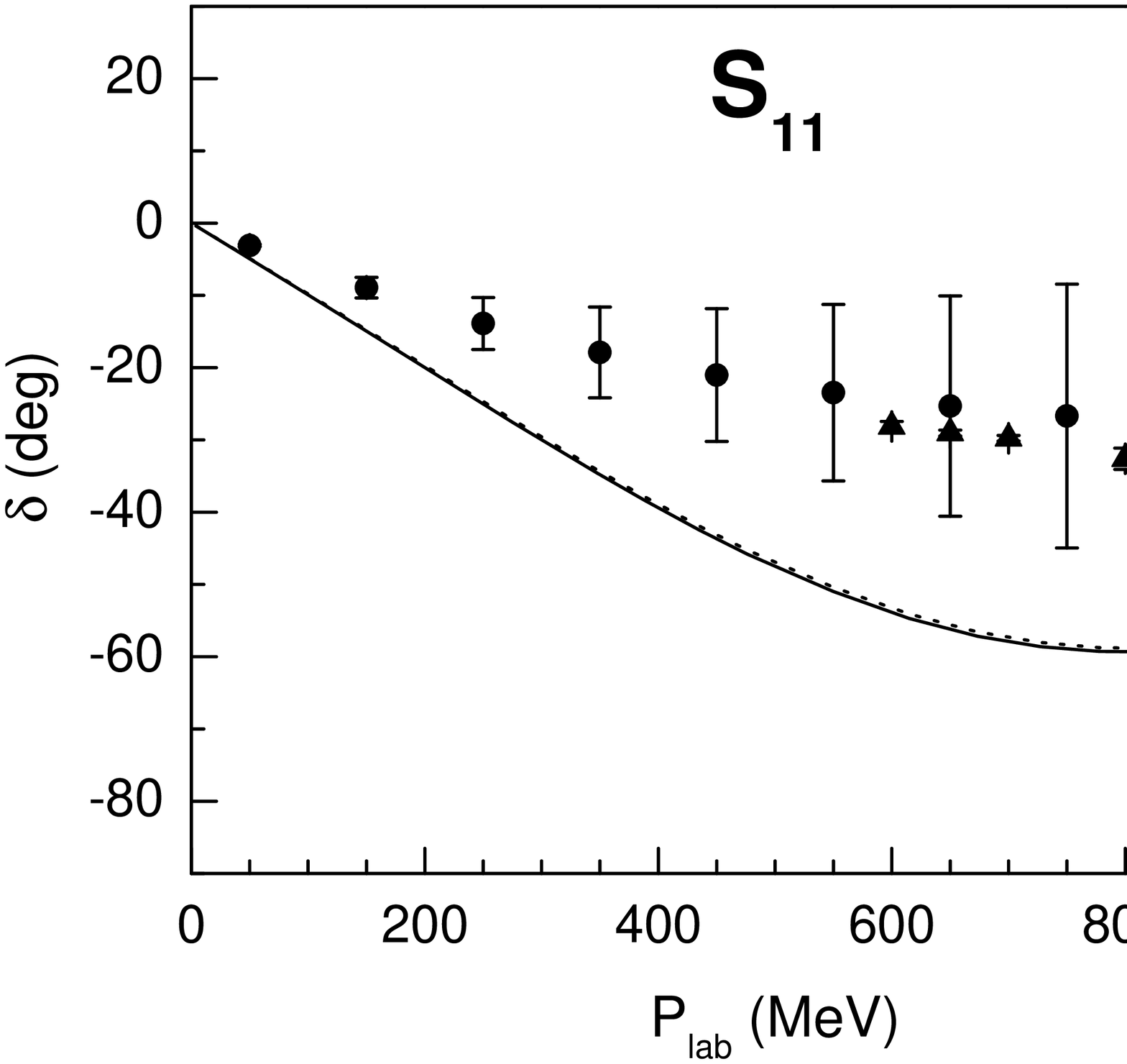,width=4.45cm}
\end{center}
\vglue -1.6cm \caption{\label{s0s1} $KN$ $S$-wave phase shifts as
a function of the laboratory momentum of kaon meson. The solid
lines show the results where $\theta^S=35^\circ$ while the dotted
lines $\theta^S=-18^\circ$. The first subscript refers to the
isospin quantum number and the second one to twice the total spin
of the channel.}
\end{figure}

\begin{figure}[h]
\begin{center}
\psfig{file=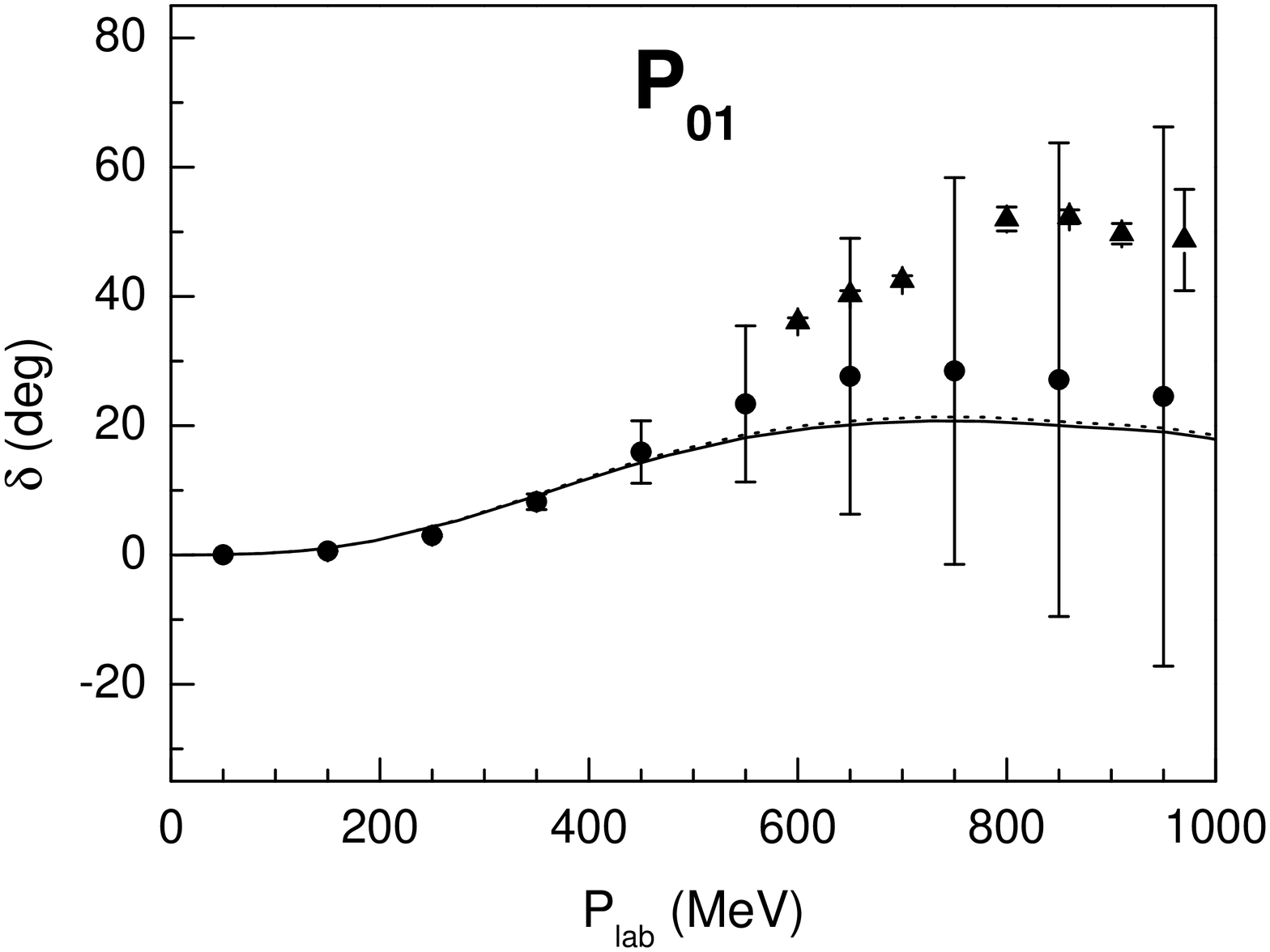,width=4.15cm}
\psfig{file=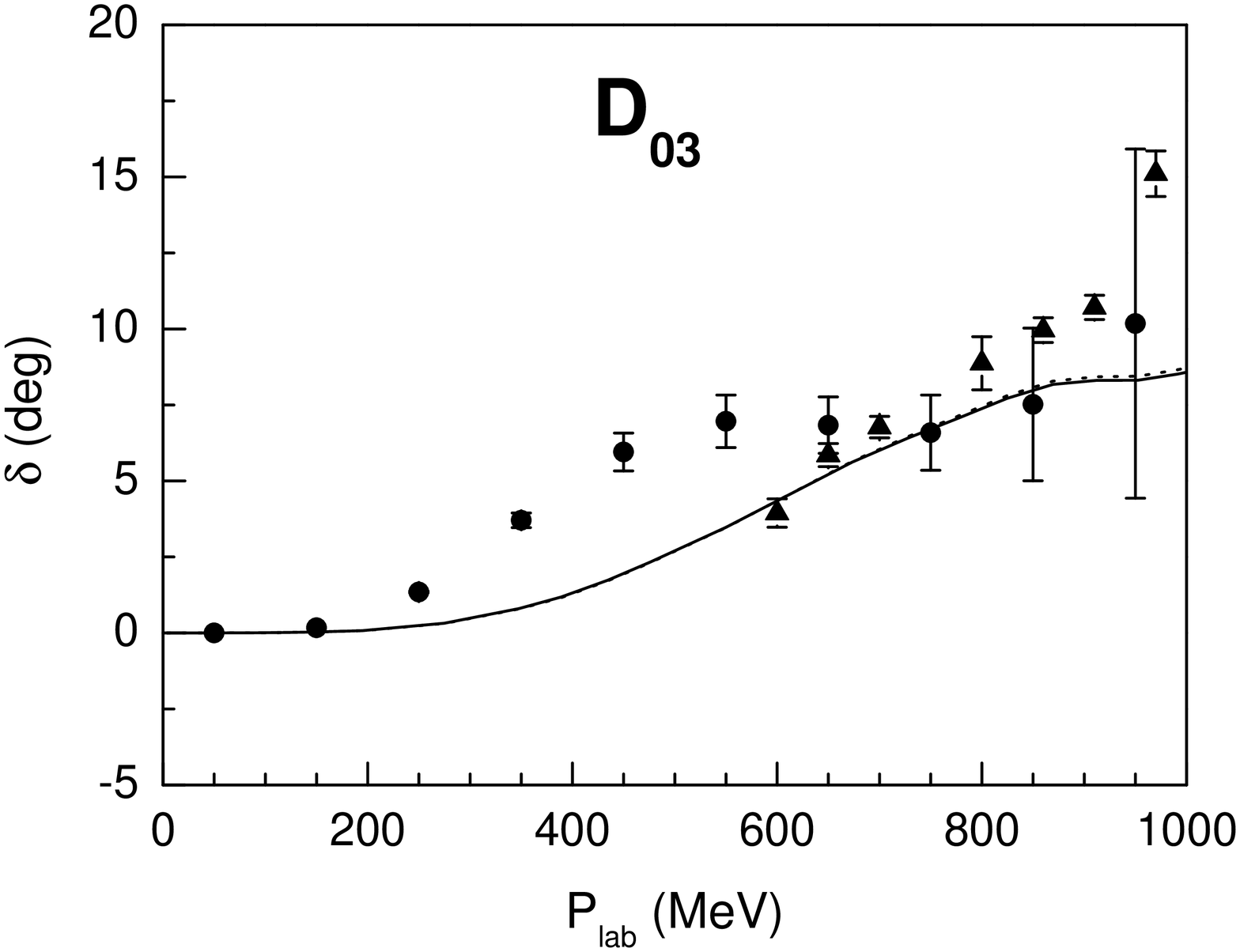,width=4.15cm}
\psfig{file=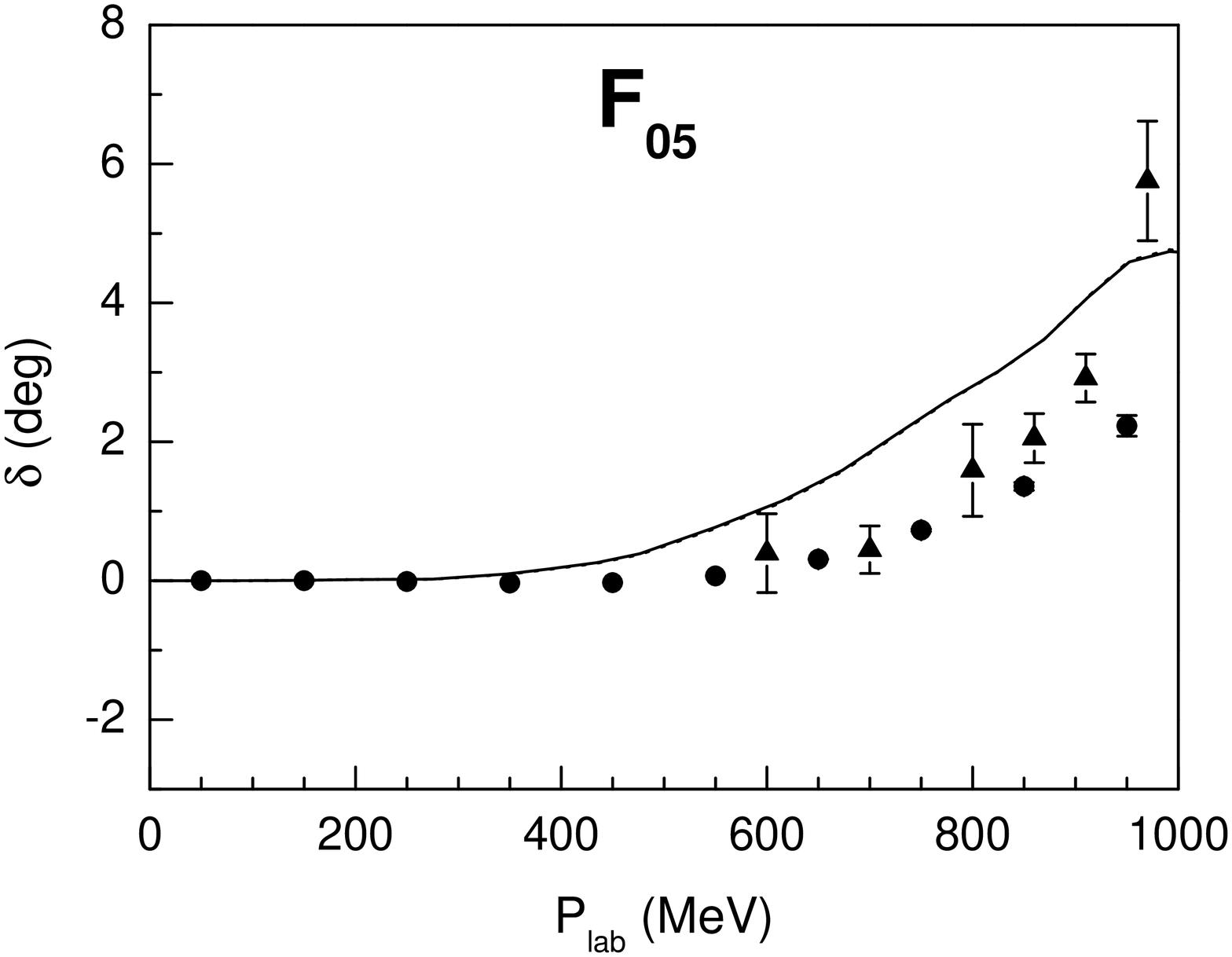,width=4.15cm}
\psfig{file=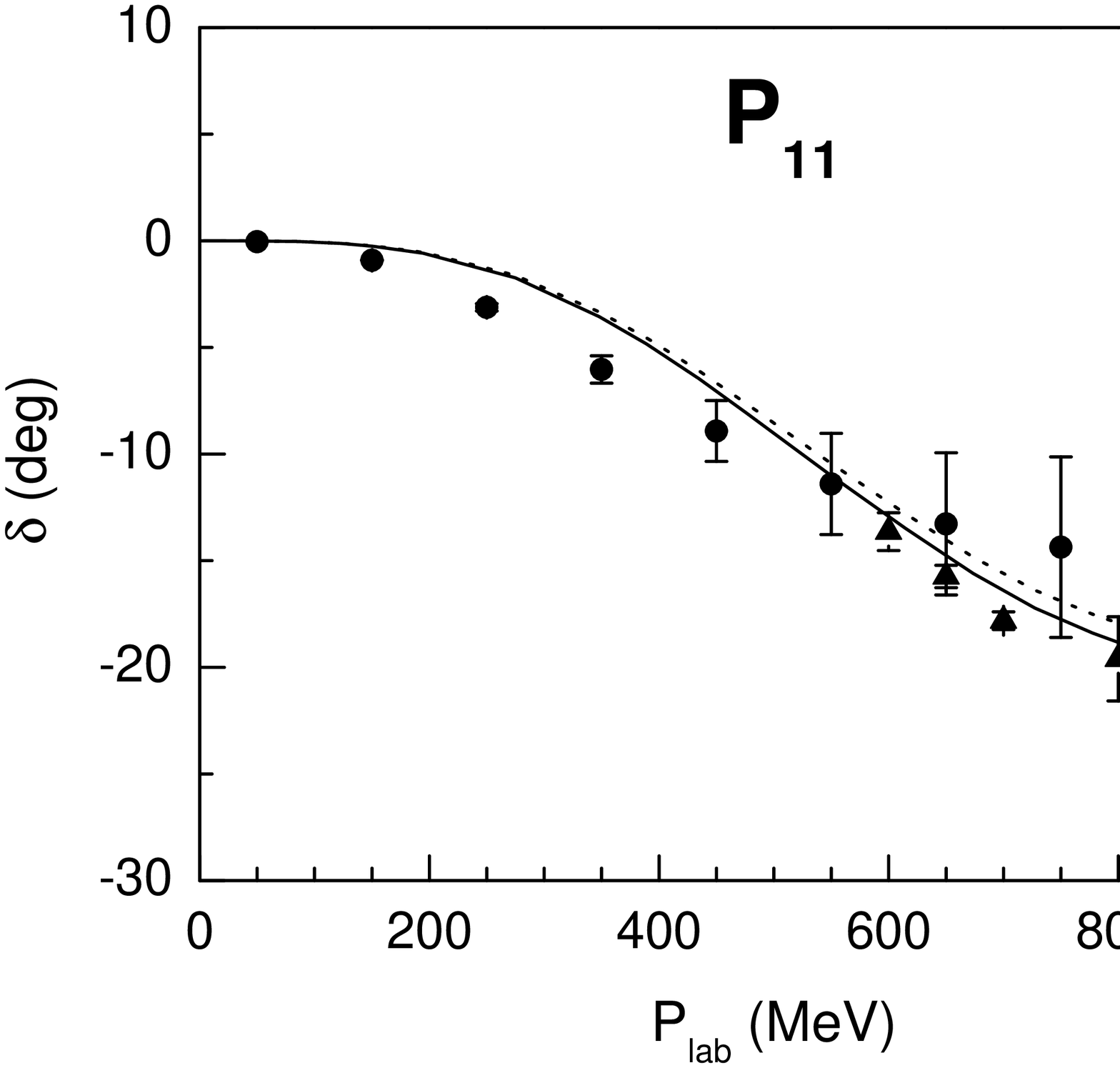,width=4.15cm}
\psfig{file=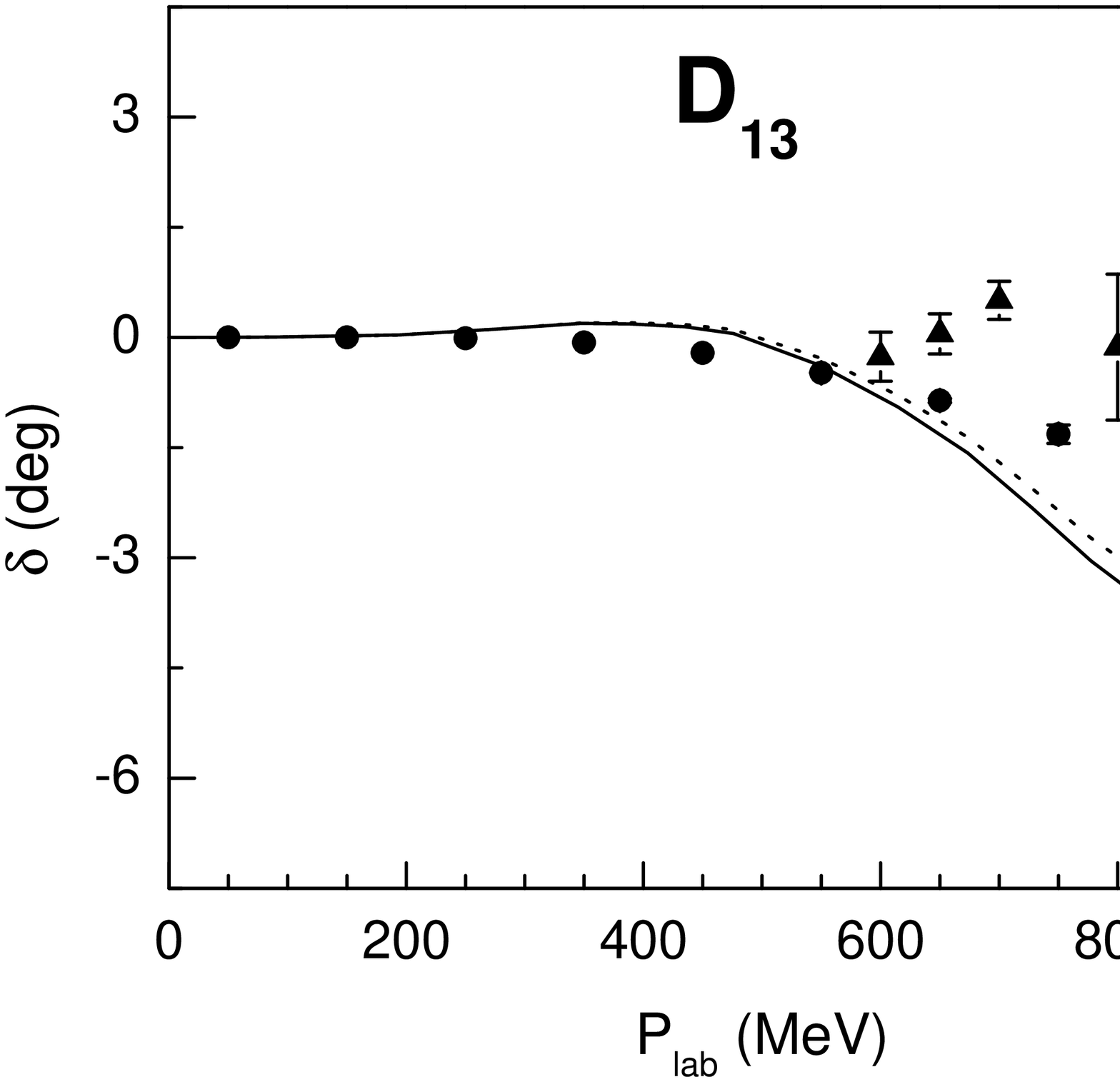,width=4.15cm}
\psfig{file=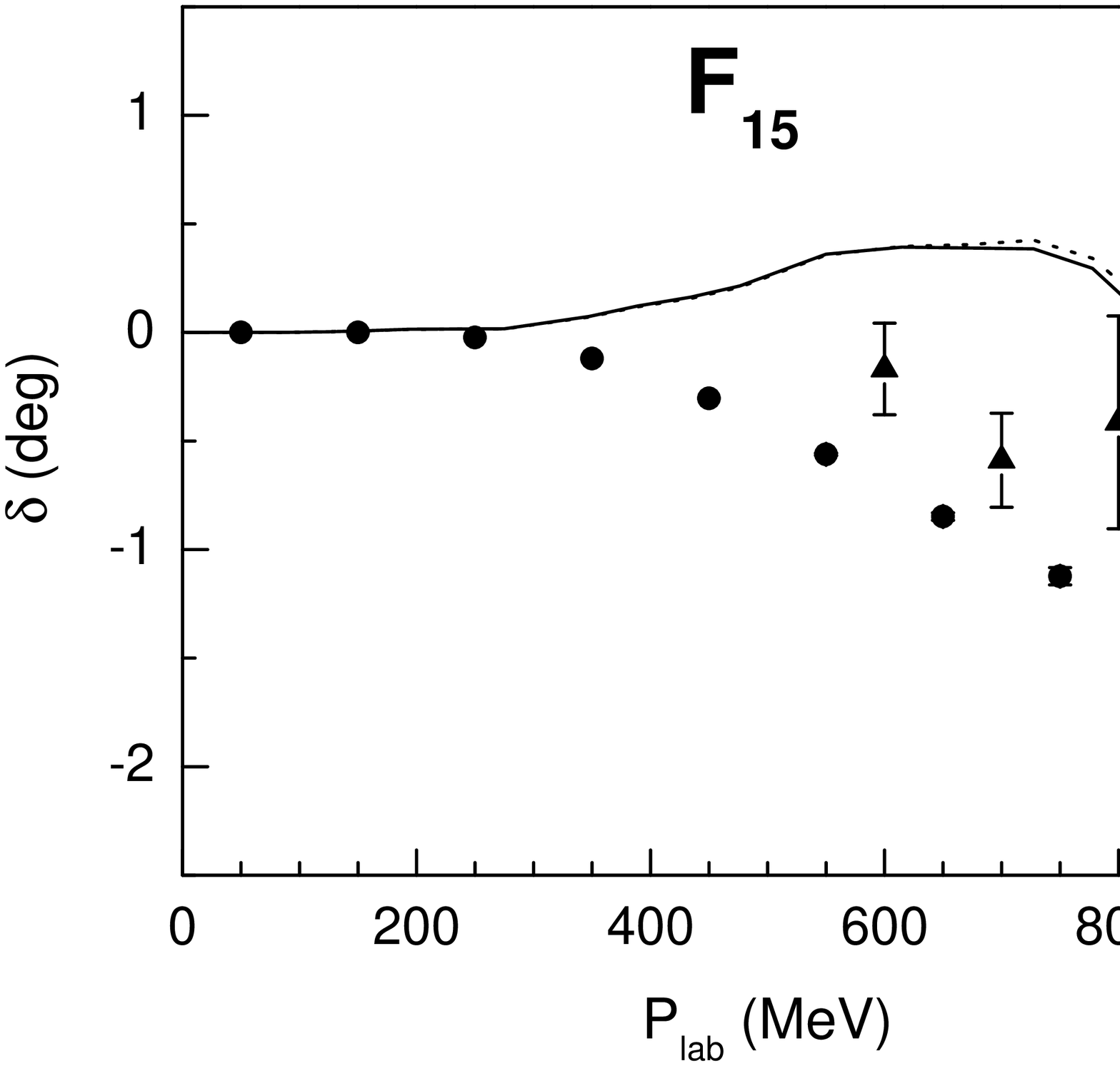,width=4.15cm}
\psfig{file=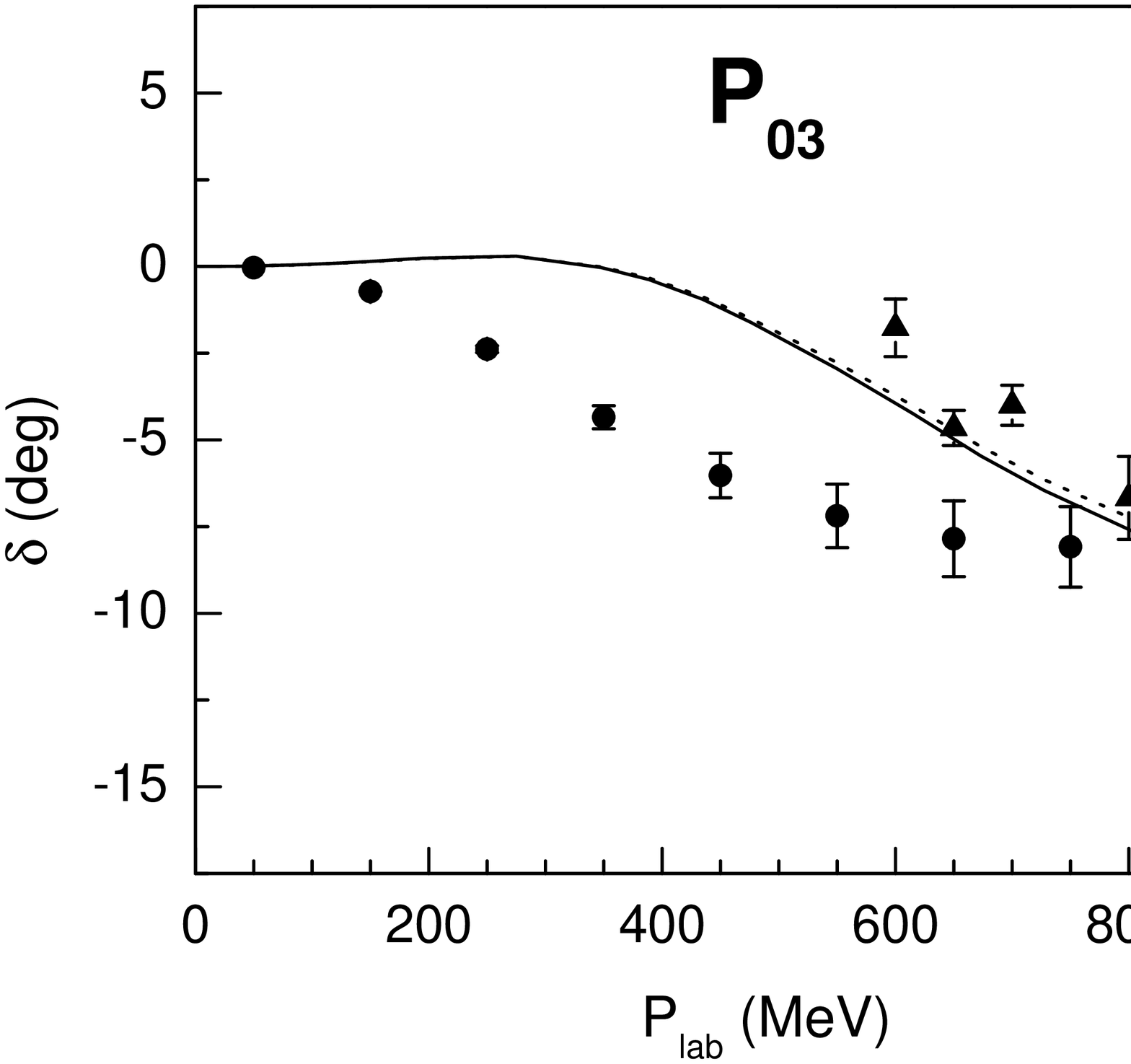,width=4.15cm}
\psfig{file=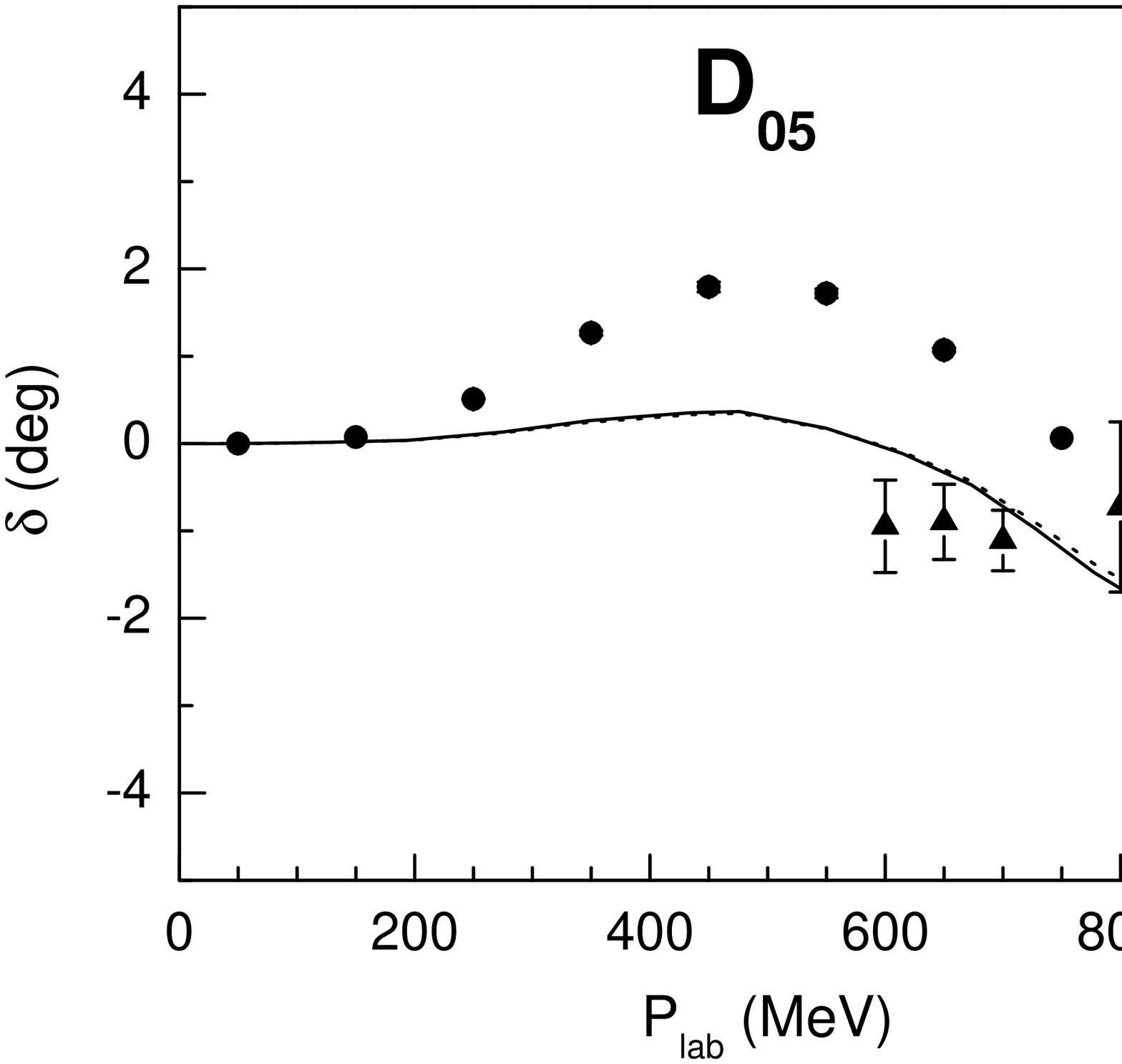,width=4.15cm}
\psfig{file=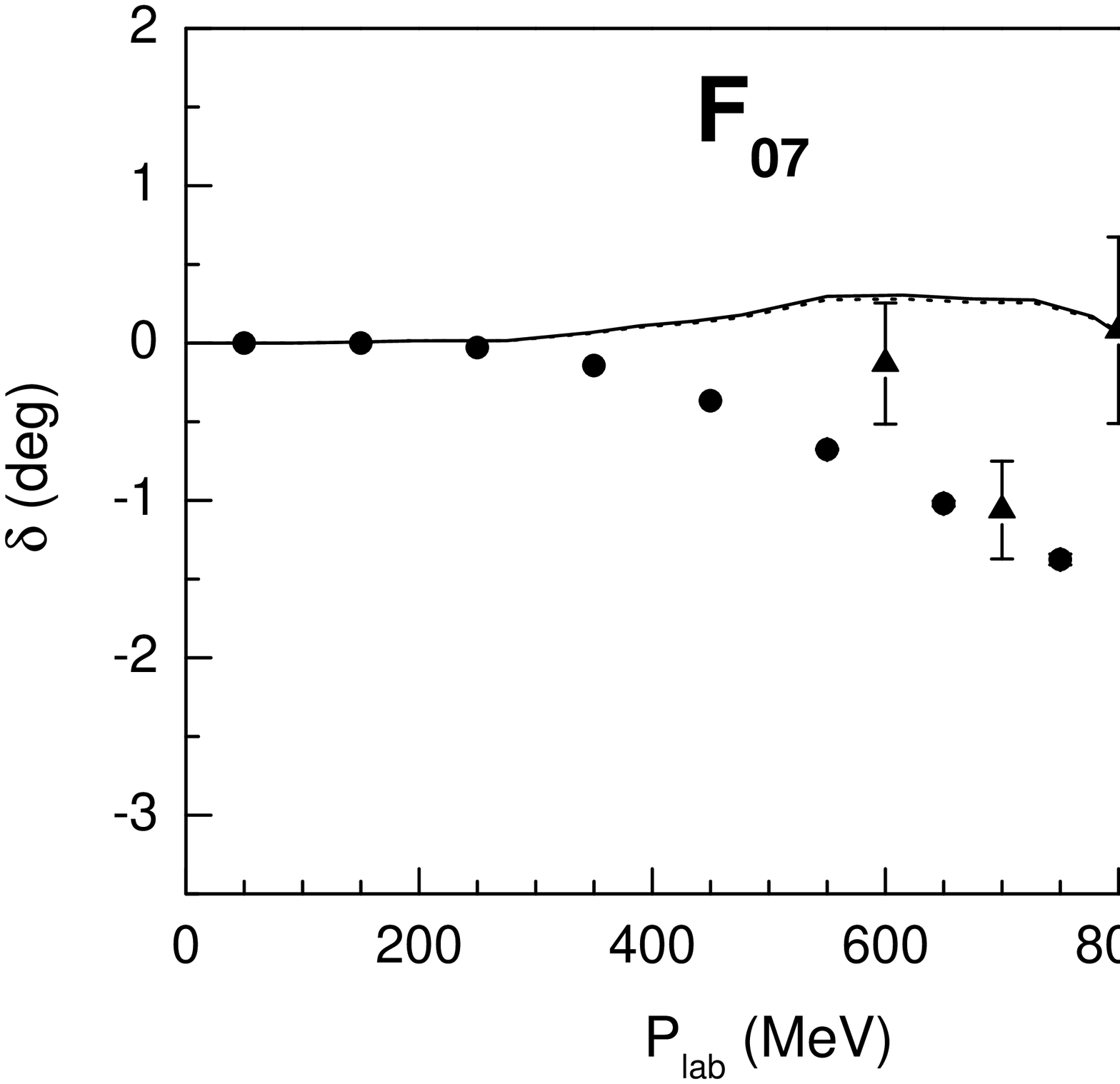,width=4.15cm}
\psfig{file=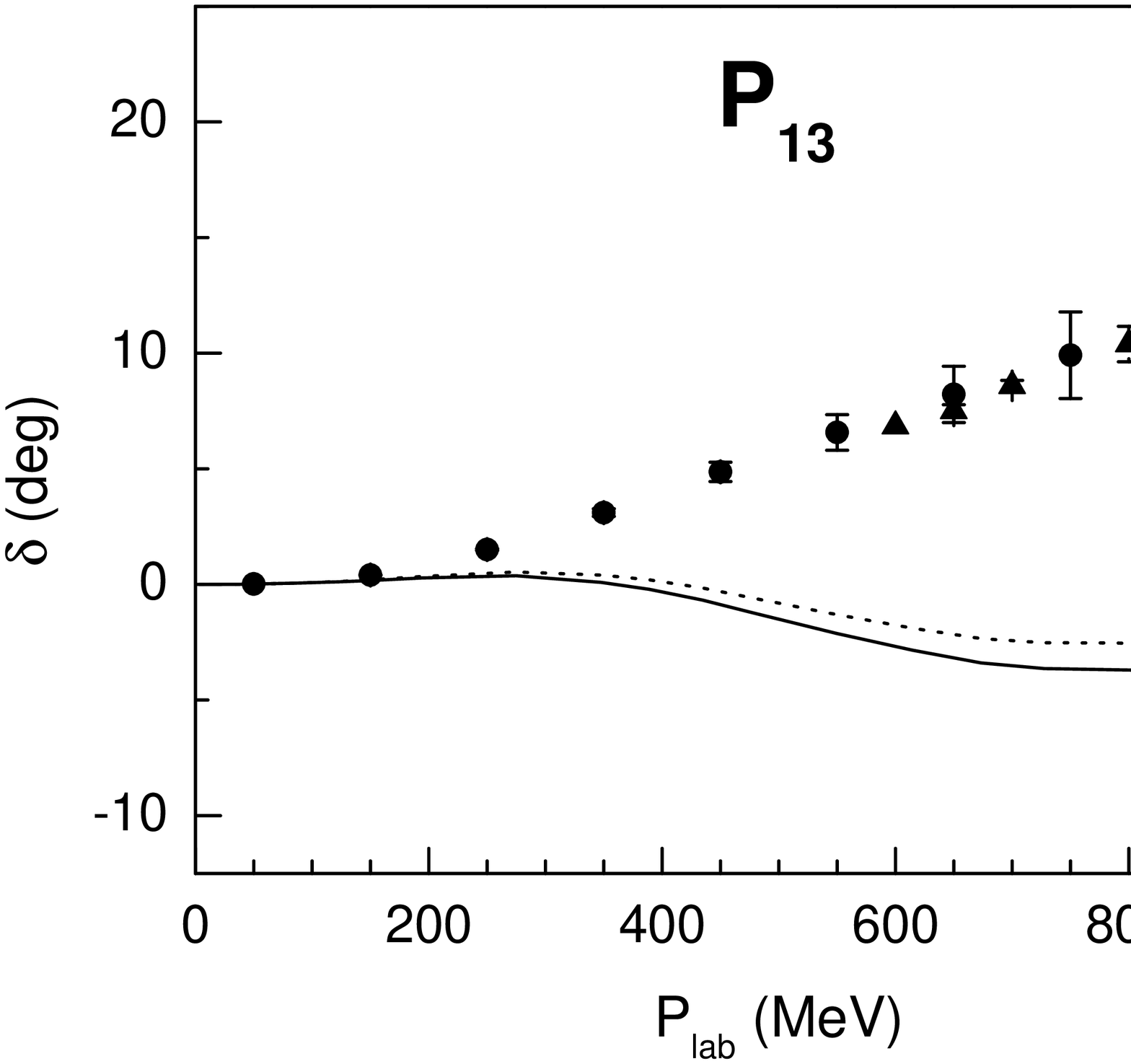,width=4.15cm}
\psfig{file=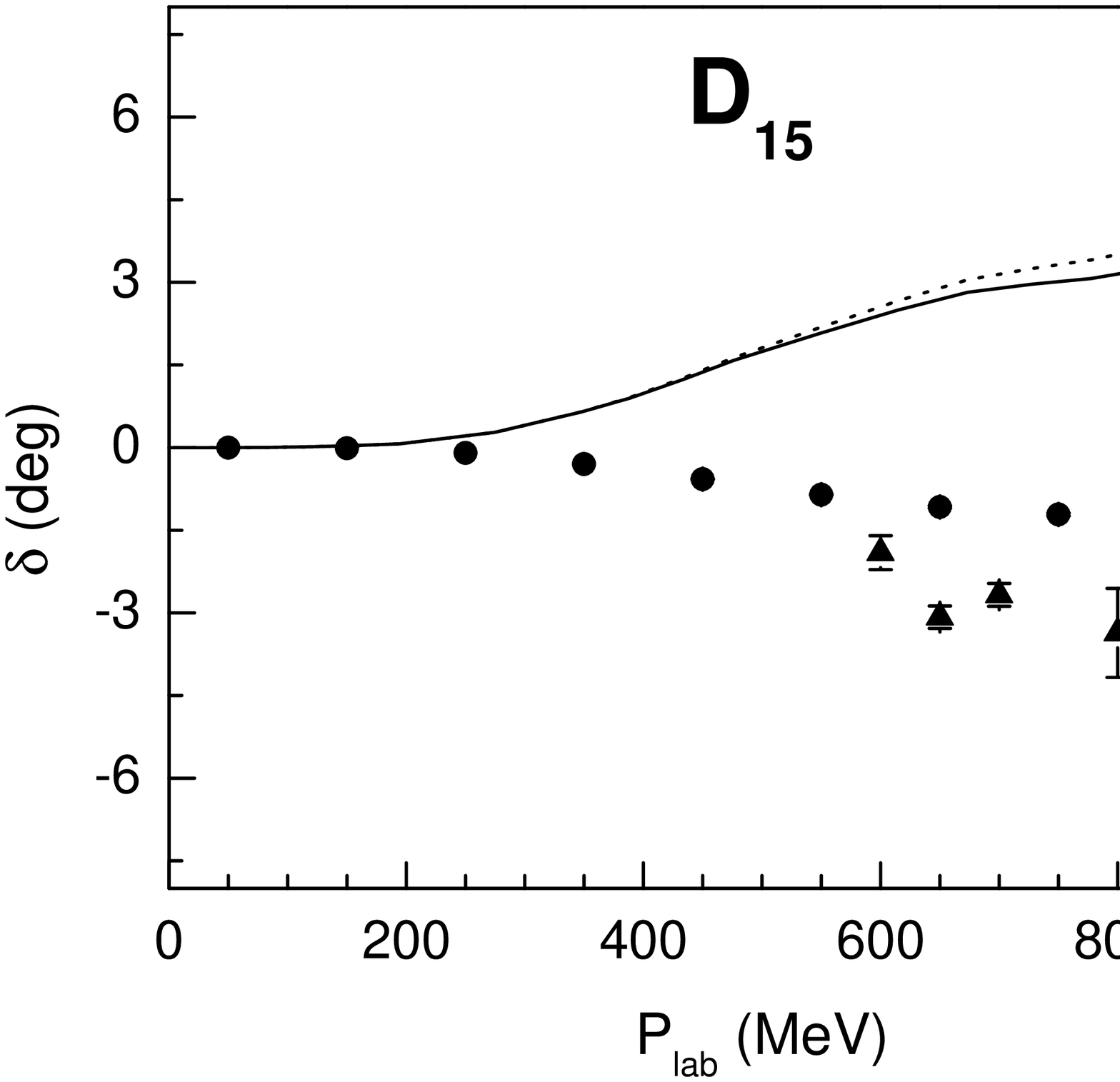,width=4.15cm}
\psfig{file=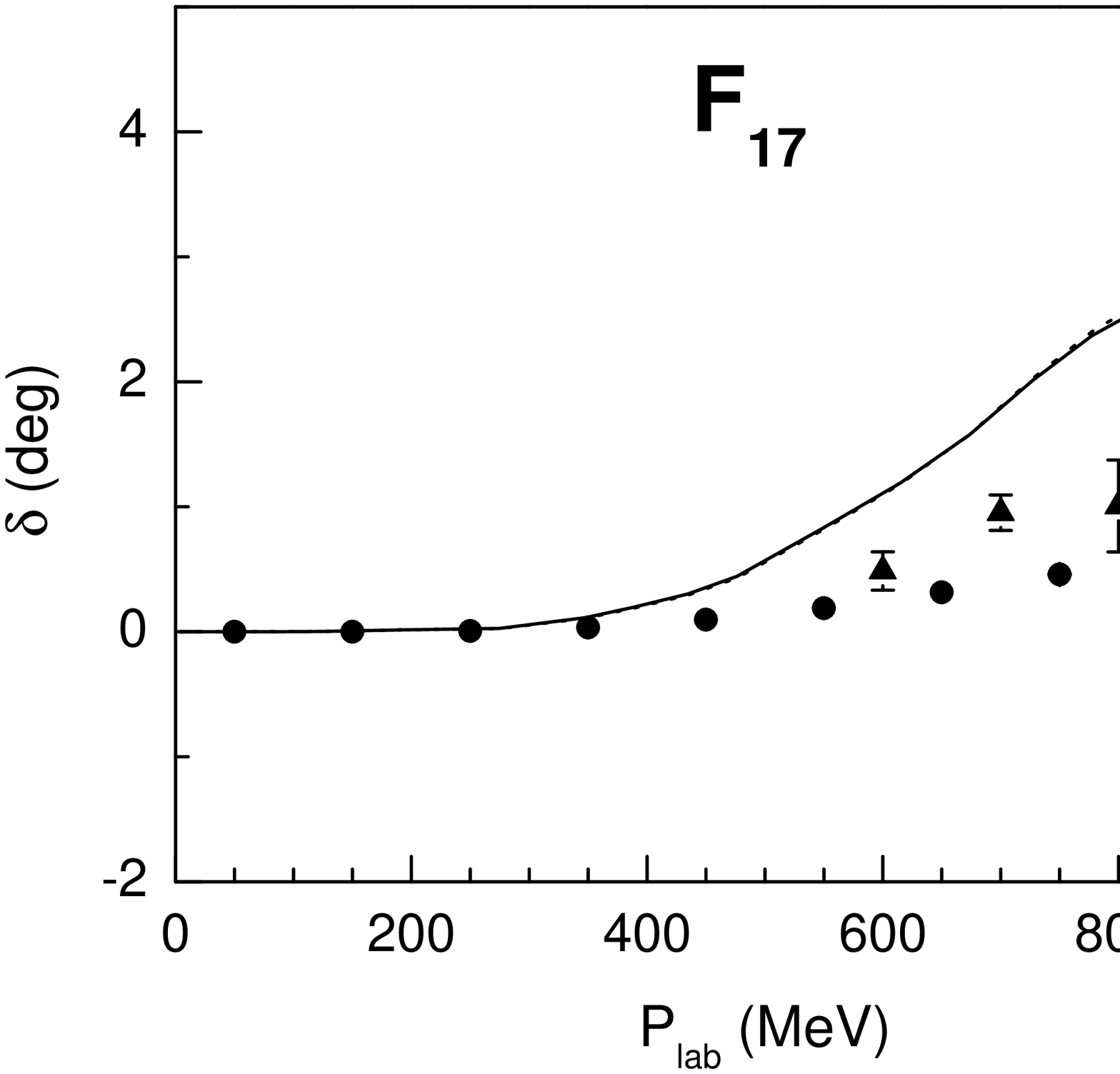,width=4.15cm}
\end{center}
\vglue -1.4cm \caption{\label{pdf} $P$, $D$, $F$ wave $KN$ phase
shifts. Same notation as in Fig. \ref{s0s1}.}
\end{figure}

A RGM dynamical calculation is made to study the $KN$ scattering
process, and the calculated phase shifts are shown in Figs.
\ref{s0s1} and \ref{pdf}. Experimental values are taken from the
analysis of Hyslop {\it et al.} \cite{jsh92} and Hashimoto
\cite{kha84}.

For the $S$-wave we obtain the correct sign of the $S_{01}$
channel phase shifts comparing with the recent RGM calculation in
which $\sigma$ and $\pi$ boson exchanges are considered
\cite{sle03}, and our results are in agreement with the
experimental data for both isospin $I=0$ and $I=1$ channels,
though for $S_{11}$ they are a little repulsive. For the higher
angular momentum results, comparing with the study of Lemaire {\it
et al.} \cite{sle03}, we now get correct signs of $P_{11}$,
$P_{03}$, $D_{13}$, $D_{05}$, $F_{15}$ and $F_{07}$ waves, and a
considerable improvement on the theoretical phase shifts in the
magnitude for $P_{01}$, $D_{03}$ and $D_{15}$ channels. We also
compare our results with those of the previous work of Black
\cite{nbl02}. Although our calculation achieves a considerable
improvement for all partial waves, the results of $P_{13}$ wave
are too repulsive in both Black's work and our present one. Maybe
the effects of the coupling to the inelastic channels and hidden
color channels should be considered in future work.

From Figs. 1 and 2 one can see the results are very similar for
the cases of $\theta^S=35^\circ$ and $\theta^S=-18^\circ$. It is
comprehensible because in both of these two cases the attraction
of $\sigma$ is reduced, just in different approaches. When
$\theta^S=35^\circ$ the interaction between $u(d)$ and $s$ quarks
vanishes, while $\theta^S=-18^\circ$ the attraction of $\sigma$
between $u$ and $d$ quarks is strongly reduced.

\section{Conclusions}

The $KN$ scattering process is studied in the chiral SU(3) quark
model, which has been extended to include an antiquark, by solving
a RGM equation. The numerical results of different partial waves
are in agreement with the experimental data except for the cases
of $P_{13}$ and $D_{15}$, which are less well described when the
laboratory momentum of the kaon meson is greater than 400 MeV in
the present investigation. It turns out that our model is
successful to be extended to study the $KN$ system, and some
useful information of the quark-quark and quark-antiquark
interactions are obtained from this study.

\section*{Acknowledgements}

This work was supported in part by the National Natural Science
Foundation of China No. 90103020.

\end{document}